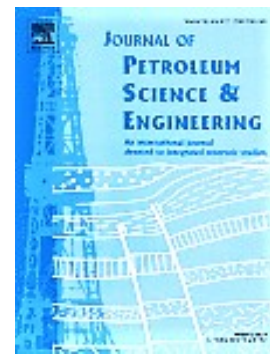

# Molecular Dynamics Studies of Interaction between Asphaltenes and Solvents

**Salah Yaseen [a] and G.Ali Mansoori [b]**

[a] Department of Chemical Engineering, University of Illinois at Chicago, Chicago, IL 60607-7052, USA; *syasee3@uic.edu*; *salahyaseen1983@gmail.com,*

[b] Departments of Bio and Chemical Engineering & Physics, University of Illinois at Chicago (M/C 063), Chicago, IL 60607-7052, USA;

*mansoori@uic.edu*; *gali.mansoori@gmail.com*

## Abstract

Understanding of the molecular interaction between asphaltenes and other molecules, which may act as its solvents, provides insights into the nature of its stability in petroleum fluids and its phase transitions. Molecular dynamics simulations were performed and reported here on systems consisting of a single asphaltene molecule and pure solvents. Three types of asphaltenes with different architectures, molecular weights, and heteroatoms content were investigated. Water and ortho-xylene were selected to be the interacting solvents. All simulations were performed by using GROMACS software. OPLS_AA potential model for hydrocarbons and SPC/E potential model for water were used in simulations. It was shown that the polar functional groups in asphaltenes were responsible for generating hydrogen bonds (HBs) between asphaltenes and water. It was also demonstrated that both electrostatic (ES) and van der Waals (vdW) interaction energies between asphaltenes and water had important roles. On the contrary, ES between asphaltenes and ortho-xylene had a minor effect as compared with the vdW. In all cases, potential energies increased rather slightly when the pressure was boosted. Moreover, they decreased noticeably when the temperature was raised. HBs between asphaltenes and water were not influenced by pressure change. Additionally, they increased slightly when the temperature was dropped.

**Keywords:** Asphaltene; molecular dynamics; solvent; ortho-xylene; water; intermolecular potential energy.

## Notations

| | |
|---|---|
| A | Asphaltene |
| COOH | carboxyl group |
| D | Debye |
| E | potential energy |
| ES | Electrostatic |
| GROMACS | GROningen MAchine for Chemical Simulations |

| H | hydrogen |
| HB | hydrogen bond |
| MD | molecular dynamics |
| NH | amine group |
| OH | hydroxyl group |
| OPLS_AA | Optimized Potential for Liquid Simulations - All Atom |
| RDF | radial distribution function |
| SPC/E | Simple Point Charge/Extended |
| vdW | van der Waals |
| W | water |

# 1. Introduction

Asphaltenes, which exist in heavy fraction of petroleum fluids and other fossil fuels, are heterocyclic macromolecules consisting fundamentally of carbon, hydrogen, and lesser amounts of other elements such as, but not limited to, sulfur, nitrogen, and oxygen, unlike hydrocarbons, the crude-oils' primary constituents, which are composed of carbon and hydrogen atoms. Besides, asphaltenes fraction of petroleum fluids may contain traces of metal atoms such as iron, nickel, and vanadium through physical association and/or encapsulation. Even though asphaltenes are a remarkable class of compounds in crude-oil, they do not have a global definition until now. However, an accepted definition of asphaltenes is by their solubility in different solvents. The more common definition is that asphaltenes are the crude-oil fraction insoluble in n-heptane and soluble in toluene. Depending on their structure, asphaltenes fall into two general architectures: Continental and archipelago architecture (Priyanto et al., 2001).

Due to asphaltenes complexity, various mechanisms describing the state of asphaltenes in crude-oils have been suggested. Solubility model, Micellar model, Fractal aggregation model, and Steric colloid model are the well-known proposed models (Park and Ali Mansoori, 1988). Naturally occurring asphaltenes in crude-oil are responsible for giving rise to harsh problems in the petroleum industry. Wetting alteration, plugging wellbore, fouling transportation pipelines, producing a stabilized oil-water emulsion, and catalyst deactivation are the primary damages caused by asphaltenes (Escobedo and Mansoori, 1997, Vazquez and Mansoori, 2000, Branco et al., 2001, Shirdel et al., 2012, Mansoori, 2009, Pacheco-Sánchez and Ali Mansoori, 2013, Vakili-Nezhaad et al., 2013, Mousavi-Dehghani et al., 2004).

MD is a computerized simulation technique, where the interaction of atoms and molecules is dictated by integrating their equations of motion. Atoms and molecules interact for a constant period, and a dynamical evolution vision is produced. In the past few years, utilization of MD simulation in the area of petroleum and natural gas systems were performed widely and successfully (Pacheco-Sánchez et al., 2003, Xue and Mansoori, 2011, Pacheco-Sánchez et al., 2004, Hu et al., 2011). Optimized Potential for Liquid Simulations – All Atom (OPLS_AA) force field was initially parameterized for biological systems. However, its applicability to asphaltenes and hydrocarbons were tested many times successfully (Xue and Mansoori, 2010, Mikami et al., 2013, Headen et al., 2009). In the OPLS-AA force field, the total interaction energy is computed as a sum of non-bonded and bonded terms of interaction energies. Non-bonded terms are expressed by electrostatic (ES) interaction energy, which is represented by Coulomb intermolecular potential, and van der Waals (vdW) interaction energy, which is represented by a Lennard-Jones intermolecular potential. Bonded terms include stretching, bending, and torsional. The total intermolecular potential energy function is expressed by Equation 1 (Jorgensen et al., 1996).

$$E_{ij} = \sum_i \sum_j \left\{ \frac{q_i q_j e^2}{r_{ij}} + 4\epsilon_{ij} \left( \left( \frac{\sigma_{ij}}{r_{ij}} \right)^{12} - \left( \frac{\sigma_{ij}}{r_{ij}} \right)^6 \right) \right\} \qquad (1)$$

In the present study, our goal was to find out how asphaltenes interact with the surrounded molecules of pure solvents. Knowing these interaction energies may give us some in-depth understanding of the behavior of asphaltenes in oil-water systems, which is of major interest in water-flooding for petroleum production and emulsion formation. We chose ortho-xylene since it is known as the best solvent and water as the worst solvent for asphaltenes. Plenty of MD simulations was implemented using an open source GROningen MAchine for Chemical Simulations (GROMACS) software (version 5.1.2). OPLS_AA force field for hydrocarbons and Simple Point Charge/Extended (SPC/E) intermolecular potential model for water molecules were employed. Our simulations were executed for the temperature range of 300 K - 360 K and the wide pressure range of 1 bar - 1000 bar. We used three model asphaltenes termed A1, A2, and A3. Two-dimensional sketches of simulated model asphaltenes are depicted in **Figure 1**. As a complement to previous studies, which concerned asphaltenes–asphaltenes interaction in different solvents, we focused on the molecular interactions between asphaltenes and other molecules. Therefore, a single asphaltene molecule was employed, which was surrounded by many molecules of ortho-xylene or water, to avoid the effect of asphaltenes–asphaltenes interaction.

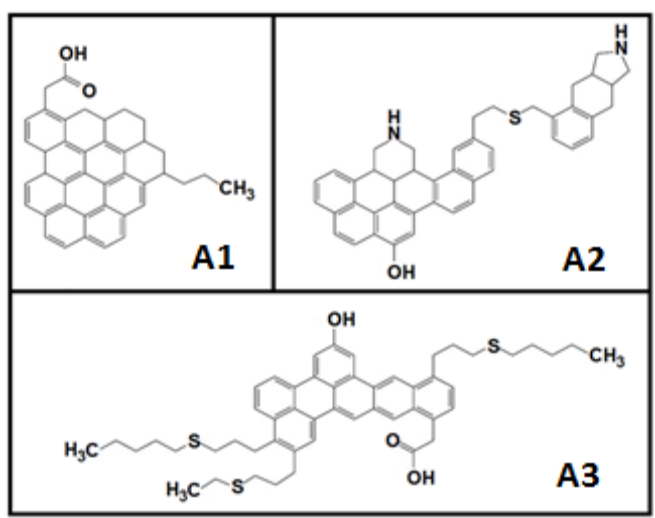

**Figure1.** Two-dimensional structure of model asphaltenes employed in this study. The structures were drawn by GChemPaint chemical structures editor (version 0.14.10-2) and further edited manually.

## 2. Material and methods

### 2.1 Molecular Models

Three asphaltenes representatives were used as model asphaltenes. The selected asphaltenes were adopted with some modifications (adding heteroatoms) from models proposed by Boek et al. (2009) (Boek et al., 2009), which was generated based on experimental data. Two typical island model asphaltenes were represented by A1 and A3 while A2 was a representative archipelago asphaltene. The detailed description of structure and composition of these asphaltenes are presented in **Figure 1** and **Table 1**. The purpose beyond the selection of three model asphaltenes was attributed to various reasons. Asphaltenes aggregation, as well as asphaltenes participating in creating a rigid oil-water interface, is strongly influenced by (i) the kind of asphaltenes whether it is archipelago or island asphaltenes (Yang et al., 2015, Kuznicki et al., 2008), (ii) the size of asphaltenes molecules (Liu et al., 2015), (iii) the number and kind of heteroatoms in asphaltenes (Zhu et al., 2015).

**Table 1.** The structure of model asphaltenes.

| Physical specifications | A1 | A2 | A3 |
|---|---|---|---|
| Chemical formula | $C_{40}H_{30}O_2$ | $C_{44}H_{40}SN_2O$ | $C_{51}H_{60}S_3O_3$ |
| Molecular weight | 542.66 | 644.86 | 817.21 |
| No. of aromatic rings | 8 | 7 | 7 |
| No. of cycloalkanes rings | 3 | 3 | 0 |
| No. of side chains | 2 | 1 | 4 |
| Aromaticity [-] | 0.7 | 0.68 | 0.55 |
| Nitrogen | - | Two secondary amine groups | - |
| Oxygen | One carboxyl group | One hydroxyl group | One carboxyl group and One hydroxyl group |
| Sulfur | - | One sulfide group | Three sulfide groups |

### 2.2 Initial Systems Configuration

Three-dimensional structures of A1, A2, A3, and ortho-xylene molecules were generated by Avogadro software (version 1.1.1) (Hanwell et al., 2012). The initial structures of these molecules were in PDB coordinates. The topology file of each molecule coordinate was generated using MKTOP script (Ribeiro et al., 2008). The skeleton of obtained topology was then verified, and errors were modified manually by adjusting the atom type, partial charge, and parameters. Two distinct systems were built. The first system composed of a single asphaltene molecule in plenty of water molecules. The second system consisted of a single asphaltene molecule in many ortho-xylene molecules. To develop an asphaltene–water system, a single asphaltene molecule was inserted in a cubic box, which could be arranged in GROMACS by using *gmx insert* command. The asphaltene molecule was centered in the box by using *gmx editconf* command and placed at least 1.0 nm from the box edge which meant that there were at least 2.0 nm between any two periodic images of a single asphaltene. So, the size of the final box depended strongly on the size of each model asphaltenes. After defining the box dimensions, water molecules were let to fill the rest of the space in the box by using *gmx solvate* command. The number of solvents molecules that filled the box was not similar and depended on asphaltene size. The number of ortho-xylene (o.xylene) molecules in A1–o.xylene, A2–o.xylene, and A3–o.xylene systems were 402, 600, 846 and the number of water molecules in A1–water, A2–water, and A3–water systems were 1395, 1854, and 2588.

## 2.3 Simulation Details

The first step in MD simulation is energy minimization. The initial structure was relaxed by using steepest descent minimization. The minimization convergence was stopped when the maximum force was below 1000 KJ/mol/nm. To bring the system to the desired temperature, MD simulation was performed under an NVT (constant number of molecules, volume and temperature) ensemble. NVT ensemble simulation lasted for 100 ps until when the temperature reached the plateau. NPT (constant number of molecules, pressure and temperature) ensemble was then employed for 500 ps to stabilize to converge the required the pressure of the system. Upon completion of NVT ensemble and NPT ensemble, where the system was well-equilibrated at the desired temperature and pressure, the system is ready to perform production MD, which persisted for 20 ns. The graphics of MD results were obtained using the VMD software (version 1.9.2) (Humphrey et al., 1996). During all simulation steps, the following settings were utilized:

- A periodic boundary condition was applied.
- A simulation time-step of 2.0 fs was used.
- Long-range electrostatic interactions were calculated using the Particle Mesh Ewald method.
- The cut-off distance for non-bonded interactions was fixed to 1.0 nm.
- The LINCS algorithm was utilized to constrain all the bond lengths in the simulation.
- A modified Berendsen thermostat (V-rescale) with a coupling constant of 0.1 ps was adopted. Parrinello-Rahman barostat was used with coupling constant of 2.0 ps.

## 2.4 Setup of Study

The plan of this research was divided into two stages. The first stage was the simulation of a single asphaltene molecule in water solvent. The range of conditions was 300–360 K for temperature and 1–1000 bar for pressure. During this stage, twenty-seven MD simulations were implemented (nine MD simulations for each model asphaltene). The second stage was the simulation of a single asphaltene molecule in the ortho-xylene solvent. The conditions and the number of MD simulations of this step were likewise the first one. In this study, fifty-four simulations were implemented.

## 2.5 Verification of MD Simulation Validation

Two MD simulations were implemented on water and ortho-xylene systems individually to examine the applicability of OPLS_AA force field to hydrocarbons and SPC/E model. An additional MD simulation was performed on the ortho-xylene–water interface. The results of these MD simulations provided quite accurate data of densities, enthalpies of vaporization, dielectric constants, and interfacial tension when compared with the available experimental data. The interfacial tension was calculated according to the method mentioned by van Buuren et al. (1993) (van Buuren et al., 1993). All simulations continued for 10 ns. The results were computed for the last 5 ns. The detailed MD results are summarized in **Table 2**.

**Table 2.** Comparison of physical properties of ortho-xylene (o.xylene) and water obtained from MD Simulation with Experimentally Determined Values. MD and experimental results are at 298 K and 1 bar unless anything else are mentioned.

| Physical Properties | Data | O.xylene | Water | O.xylene - Water |
|---|---|---|---|---|
| Density [kg/m$^3$] | Literature | 875.53[a] | 997.13[b] | - |
| | MD | 875 | 999 | - |
| Enthalpy of vaporization [kJ/mol] | Literature | 45[c] | 43.990[d] | - |
| | MD | 44 | 46.8 | - |
| Dielectric constant [-] | Literature | 2.379[e], [f] | 77.166[g], [h] | - |
| | MD | 1.5[f] | 73.9[h] | - |
| Interfacial tension [mN/m] | Literature | - | - | 36.3[i] |
| | MD | -- | - | 37.2 |

[a](Al-Kandary et al., 2006),[b](Tanaka et al., 2001), [c](Pitzer and Scott, 1943), [d](Chickos and Acree Jr, 2003), [e](Skinner et al., 1968), [f] 300 K and 1 bar, [g](Anderson et al., 2000), [h] 302 K and 1 bar, [i](Demond and Lindner, 1993).

## 3. Results and Discussion

### 3.1 Hydrogen Bonds Between Asphaltenes and Solvents Molecules

Hydrogen bonds (HBs) is a physical bond (weaker than covalent and ionic bonds) formed between an electropositive atom (usually hydrogen) and a strongly electronegative atom (mostly oxygen or nitrogen) which has a at least one lone pair of electrons that may participate to form a hydrogen bond (Desiraju and Steiner, 2001). In this study, HBs between model asphaltenes and water molecules were formed (see **Figure 2**). Water is considered as a typical polar compound. Model asphaltenes contain various polar functional groups, which can connect with water molecules through HBs. A1 includes one carboxyl functional group, A2 contains two secondary amine and one hydroxyl functional group, and A3 contains one carboxyl and one hydroxyl functional group. A carboxyl group is supposed to be the strongest polar group in the current study because it is both a hydrogen bond acceptor (the carbonyl) and a hydrogen bond donor (the hydroxyl). Both hydroxyl and secondary amine groups are electron donor group, and they are presumed as strong polar functional groups. The polar functional groups that exist in model asphaltenes were responsible for creating HBs with water molecules. The average number of HBs between model asphaltenes and water molecules at various temperatures and pressures are reported in **Table 3**. The number of these HBs followed the order:

$$(A2\text{-}W)^{Av.\ No.\ HBs} \; > \; (A3\text{-}W)^{Av.\ No.\ HBs} \; > \; (A1\text{-}W)^{Av.\ No.\ HBs} \qquad (2)$$

$(A2\text{-}W)^{Av.\ No.\ HBs}$ was 150% and 50% larger than $(A1\text{-}W)^{Av.\ No.\ HBs}$, and $(A3\text{-}W)^{Av.\ No.\ HBs}$ respectively. This trend was because A2 has a greater number of polar functional groups than A1 and A3. The interaction between asphaltenes and ortho-xylene molecules was entirely different from the interaction between asphaltenes and water molecules. In asphaltene–o.xylene system, HBs were not formed because the ortho-xylene structure does not contain nitrogen and oxygen.

**Table 3.** Average number of HBs between a model asphaltene molecule and water molecules at various temperatures and pressure.

| T [K] | P [bar] | $(A1\text{-}W)^{Av.\ No.\ HBs}$ | $(A2\text{-}W)^{Av.\ No.\ HBs}$ | $(A3\text{-}W)^{Av.\ No.\ HBs}$ |
|---|---|---|---|---|
| 300 | 1 | 2.25 | 5.78 | 3.90 |
|  | 100 | 2.25 | 5.80 | 3.90 |
|  | 1000 | 2.25 | 5.82 | 3 .91 |
| 330 | 1 | 2.23 | 5.62 | 3.84 |
|  | 100 | 2.23 | 5.65 | 3.85 |
|  | 1000 | 2.23 | 5.69 | 3.87 |
| 360 | 1 | 2.20 | 5.33 | 3.68 |
|  | 100 | 2.21 | 5.37 | 3.71 |
|  | 1000 | 2.21 | 5.44 | 3.76 |

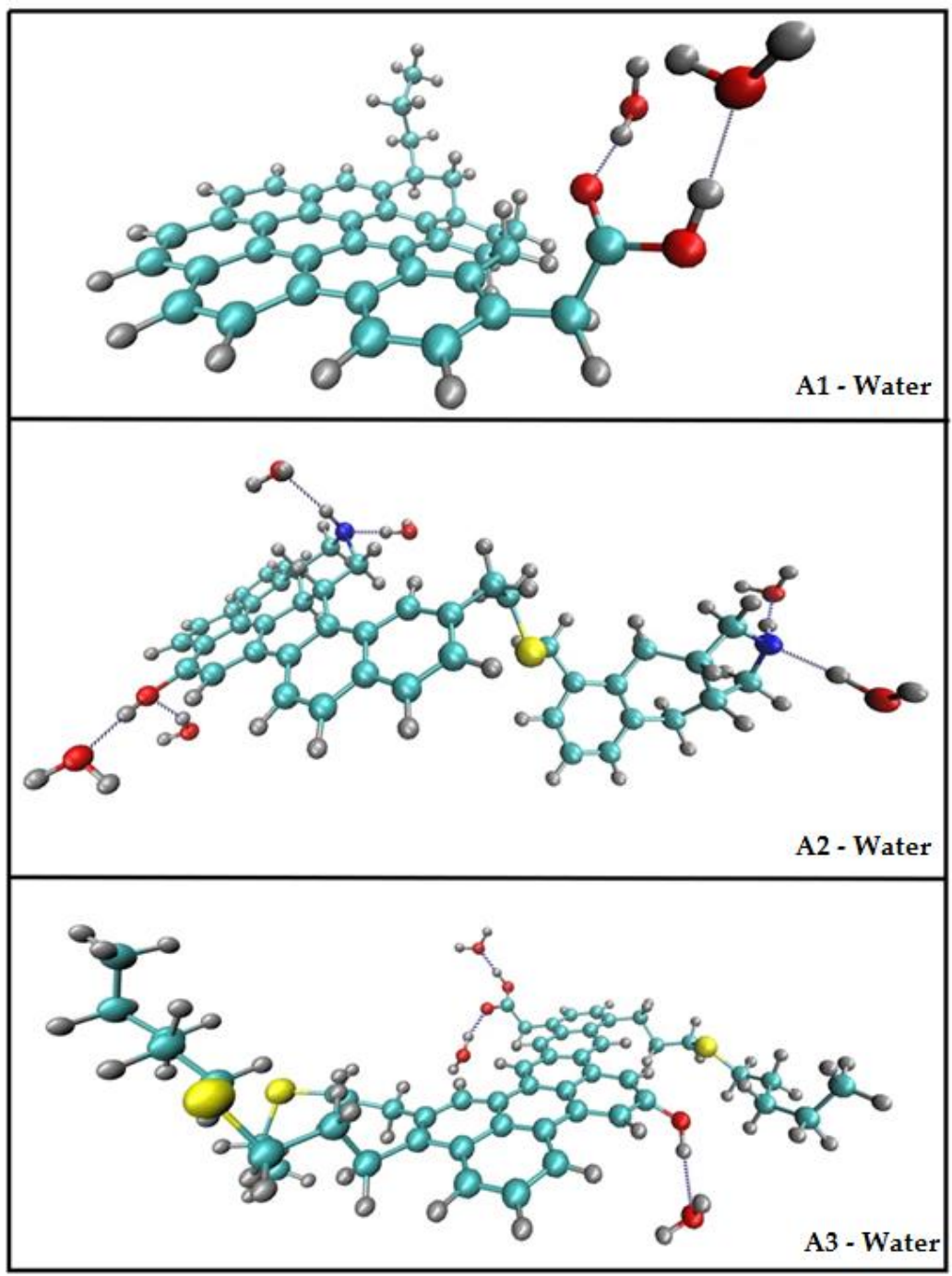

**Figure 2.** Three-dimensional structure snapshot of model asphaltene–water interaction. HBs formed between polar functional groups and water are represented by dashed lines. Color code: water molecules are in red (oxygen) and gray (hydrogen). Model asphaltenes are represented by cyan (carbon), gray (hydrogen), blue (nitrogen), yellow (sulfur), and red (oxygen). In this figure, for clarification purpose, all other water molecules are not shown, except the ones that form HBs with model asphaltenes.

### *3.2 Radial Distribution Functions (RDFs) and Structural Analysis of Hydrogen Bonds*

Radial Distribution Functions (RDFs) may be employed to identify HBs between model asphaltenes and water. RDFs of hydrogen and oxygen in water molecules around hydrogen, nitrogen, and oxygen of polar functional groups of asphaltenes are shown in **Figures 3** and **4**. A comparison of these figures indicated that HBs between the hydrogen atoms of polar functional groups and oxygen atoms of water molecules were stronger than all the other HBs. Moreover, RDFs pointed out that HBs between $H_{COOH}$ and $O_{water}$ in A1 and A3 were the strongest (see **Figure 3a**). In return, RDFs confirmed that HBs between $O_{CO\underline{O}H}$ - $H_{water}$ in A1 and A3 were weaker than all the other HBs (see **Figure 4a**). A shoulder was observed in the range of 0.19–0.25 nm. This shoulder means that $H_{water}$ atoms around $O_{CO\underline{O}H}$ decreased in the range 0.19–0.25 nm so that there was little possibility of forming HBs within this range. The peak at 0.32 nm was due to the presence of water molecules in this location which formed strong HBs with $H_{COO\underline{H}}$ atom.

Further, the structure of HBs between the carboxyl group and water was analyzed using RDFs as depicted in **Figure 5**. Based on **Figures 3a** and **4b**, the possibility of generating HBs of $H_{COO\underline{H}}$ - $O_{water}$ (r = 0.165 nm) and $O_{C\underline{O}OH}$–$H_{water}$ (r = 0.18 nm) was strong. This result is in good agreement with the findings of Jian et al. (2014) who worked on the aggregation of asphaltenes in toluene and water. They found that the mean distance of donor-acceptor of HBs (oxygen of carbonyl of model asphaltenes–hydrogen of water) was about 0.18 nm (Jian et al., 2014). Moreover, based on **Figure 4a**, the possibility of generating HBs of $O_{CO\underline{O}H}$ – $H_{water}$ was little at r = 0.225 nm (shoulder of $O_{CO\underline{O}H}$ – $H_{water}$) or r = 0.32 nm (peaks of $O_{CO\underline{O}H}$ – $H_{water}$). A carboxyl group comprises of two moieties, i.e., hydroxyl (OH) and carbonyl (C=O). Three atoms of a carboxyl group may involve in HBs interactions with water molecules, which are one hydrogen and two oxygen atoms, as depicted in **Figure 5**. RDFs indicated that HBs formed between the hydroxyl of the carboxyl group and water were stronger than the ones formed between the hydroxyl of alcohols and water (see **Figures 3a**, **3b**, **4a**, and **4c**). This behavior was because the hydroxyl of the carboxyl group is more strongly polarized than the hydroxyl of alcohols (stronger dipole moment) due to the existence of the adjacent carbonyl moiety.

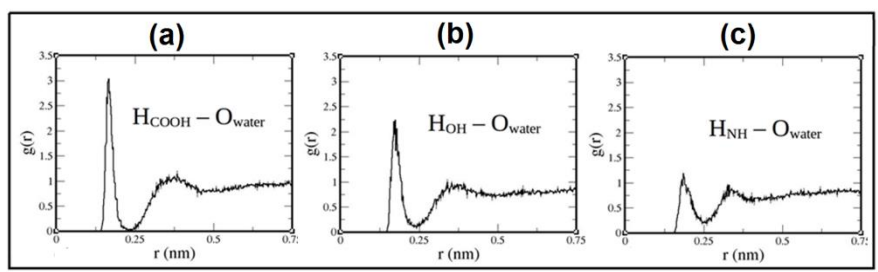

**Figure 3.** The radial distribution function between hydrogen atoms of polar functional groups of model asphaltenes and oxygen atoms of water at 300 K and 1 bar.

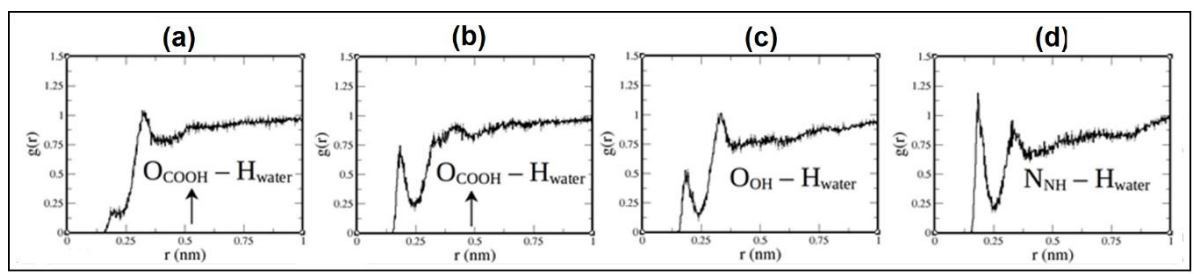

**Figure 4.** The radial distribution function between heteroatoms of polar functional groups of model asphaltenes and hydrogen atoms of water at 300 K and 1 bar.

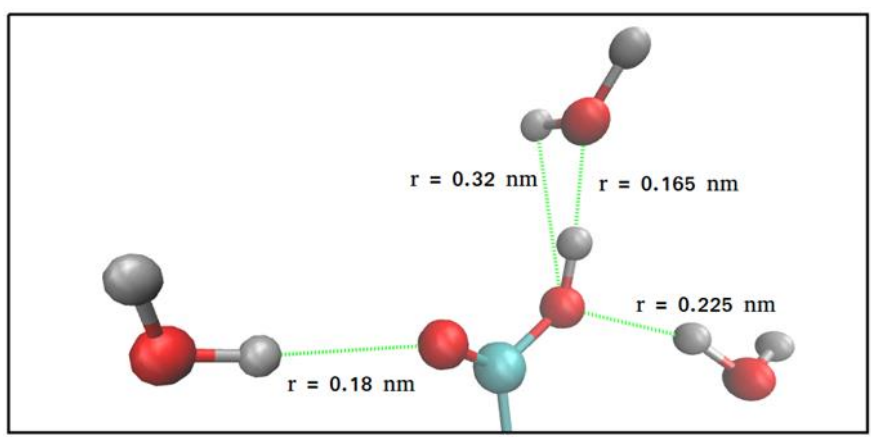

**Figure 5.** The structure of HBs between COOH group of asphaltenes and water at 300 K and 1 bar. HBs formed between polar functional groups and water are represented by dashed green lines. Color code: water molecules are in red (oxygen) and gray (hydrogen). COOH group is represented in cyan (carbon), gray (hydrogen), and red (oxygen). In this figure for clarification purpose, all other water molecules are not shown, except the ones that form HBs with model asphaltenes.

### 3.3 Interaction Potentials Between Model Asphaltenes and Solvent Molecules

In addition to HBs, two kinds of non-bonding interaction, namely electrostatic (ES) and van der Waals (vdW), were investigated. Van der Waals potential energy ($E_{VDW}$) varies mainly by two factors, specifically, surface area (geometry of molecule) and electronic polarizability (molecular size). The larger the surface area of the molecule, the larger chances of more vdW interactions. Polarizability measures the ease of distorting an electron cloud around an atom as a response to changes in its electronic environment. Polarizability of an atom is affected by the number of electrons and the radius of an atom. Large molecules are usually associated with greater polarizability. They are more polarizable than smaller ones because there are more electrons to deform. Electrostatic potential energy ($E_{ES}$) is influenced by the polarity of interacted molecules which could be expressed by a dipole moment. The greater the difference in electronegativity values of bonded atoms, the larger the dipole moment. The existence of heteroatoms plays a significant role in boosting molecule polarity (Tipler and Mosca, 2007).

**Table 4** compares contributions of potential energies between model asphaltenes and their solvents. It presents four distinct potential energies, i.e., $E_{VDW}$ (asphaltene–water), $E_{ES}$ (asphaltene–water), $E_{VDW}$ (asphaltene–o.xylene), and $E_{ES}$ (asphaltene–o.xylene). Results indicated that all these potential energies were strong attractive interactions except the case of ES between asphaltenes and ortho-xylene molecules which showed weak attractive interaction. $E_{VDW}$ (asphaltene–water) and $E_{VDW}$ (asphaltene–o.xylene) were fundamentally due to the large size of model asphaltenes which means more space for electron distribution, and thus, more possibilities for an instantaneous dipole moment. $E_{ES}$ (asphaltene–water) could be attributed to the high polarity of model asphaltenes and water. The Existence of heteroatoms in the structure of asphaltenes is responsible for their high polarity. Low polarity of ortho-xylene (lack to polar functional groups) resulted in weak ES attractive interaction between asphaltenes and ortho-xylene molecules.

For all model asphaltenes, $E_{ES}$ (asphaltene–water) was higher than $E_{ES}$ (asphaltene–o.xylene). For example, at 300 K and 1 bar, $E_{ES}$ (A3–water) was -182 KJ/mol. At the same condition, $E_{ES}$ (A3–o.xylene) was -43. This noticeable gap between potential energies was due to the large difference in polarity between water and ortho-xylene. Water has a dipole moment of 1.87 D while the dipole moment of ortho-xylene equals to 0.64 D. For all model asphaltenes, $E_{VDW}$ between asphaltenes and ortho-xylene molecules was stronger than $E_{VDW}$ between asphaltenes and water molecules. For instance, at 300 K and 1 bar, $E_{VDW}$ (A1–water) was -178 KJ/mol while $E_{VDW}$ (A1–o.xylene) was -271 KJ/mol. This trend was because the ortho-xylene molecule has a larger surface area than the water molecule (see **Figure 6**).

In asphaltene–o.xylene system, $E_{VDW}$ was higher than $E_{ES}$ for all model asphaltenes (see **Table 4**) . This behavior was because of the low polarity of ortho-xylene, which resulted in weak ES attractive interaction as we discussed earlier. On the contrary, the interaction between asphaltenes and water in asphaltene–water system showed a different approach. In the case of A1 and A3, $E_{ES}$ was lower than $E_{VDW}$. In the case of A2, $E_{ES}$ was higher than $E_{VDW}$ (see **Table 4**). So, as asphaltenes have more polar functional groups, stronger ES interaction between asphaltenes and water is expected.

**Table 4.** Potential energies [KJ/mol] between model asphaltenes (A) and water (W) and ortho-xylene (X) at various temperatures and pressures.

| T [K] | P [bar] | Asphaltene | $E_{ES}$ ($A_i$-W) | $E_{VDW}$ ($A_i$-W) | $E_{ES}$ ($A_i$-X) | $E_{VDW}$ ($A_i$-X) |
|---|---|---|---|---|---|---|
| 300 | 1 | A1 | -103 | -178 | -21 | -271 |
| | | A2 | -244 | -186 | -33 | -316 |
| | | A3 | -182 | -269 | -43 | -439 |
| | 100 | A1 | -103 | -179 | -22 | -274 |
| | | A2 | -245 | -186 | -33 | -317 |
| | | A3 | -183 | -271 | -43 | -441 |
| | 1000 | A1 | -104 | -187 | -23 | -287 |
| | | A2 | -245 | -188 | -37 | -332 |
| | | A3 | -184 | -285 | -46 | -466 |
| 330 | 1 | A1 | -100 | -170 | -19 | -258 |
| | | A2 | -234 | -178 | -29 | -301 |
| | | A3 | -176 | -250 | -37 | -412 |
| 360 | 1 | A1 | -96 | -161 | -17 | -247 |
| | | A2 | -220 | -172 | -25 | -284 |
| | | A3 | -170 | -237 | -33 | -391 |

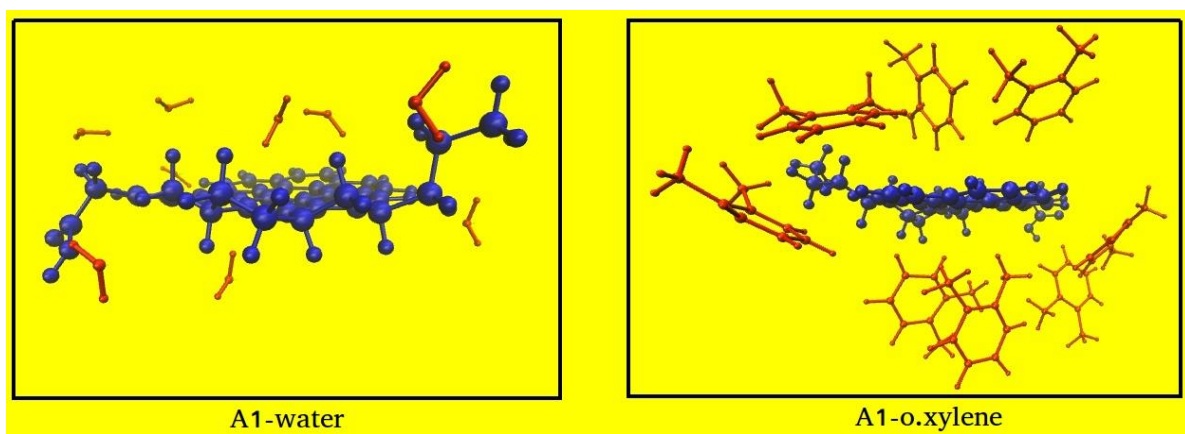

**Figure 6.** snapshots of A1 surrounded by the "first-nearest-neighbor" layer water and ortho-xylene (o.xylene) molecules at 300 K and 1 bar. Color code: A1 is blue, water is red, and ortho-xylene is red. In this figure, we only show 9 of 1395 molecules of water and 8 of 402 molecules of ortho-xylene used in simulation, which are the nearest-neighbors to asphaltene.

$E_{VDW}$ between asphaltene and solvents molecules followed the arrangement (see **Table 4**):

$$E_{VDW} (A3–o.xylene) > E_{VDW} (A2–o.xylene) > E_{VDW} (A1–o.xylene) \tag{3}$$

$$E_{VDW} \text{ (A3–water)} > E_{VDW} \text{ (A2–water)} > E_{VDW} \text{ (A1–water)} \qquad (4)$$

This behavior could be attributed to the size of asphaltenes molecules which follows the order (see **Figure 1**):

$$\text{size of A3 molecule} > \text{size of A2 molecule} > \text{size of A1 molecule} \qquad (5)$$

$E_{ES}$ of asphaltene–water followed the trend (see **Table 4**):

$$E_{ES} \text{ of A2–water} > E_{ES} \text{ of A3–water} > E_{ES} \text{ of A1–water} \qquad (6)$$

For example, at 300 K and 1 bar, $E_{ES}$ of A1–Water, $E_{ES}$ of A2–water, and $E_{ES}$ of A3–water were -103 KJ/mol, -244 KJ/mol, and -182 KJ/mol respectively. This behavior was due to the number of polar functional groups (nitrogen and oxygen) in model asphaltenes. The higher number of polar functional groups, the higher $E_{ES}$ of asphaltene- water.

### 3.4 The Effect of Temperature and Pressure Changes on Molecular Interaction

Results indicated that temperature and pressure changes influenced $E_{ES}$ and $E_{VDW}$ of asphaltene–water and asphaltene–o.xylene. Lowering of interaction energies was observed when the temperature was increased, or pressure was decreased as reported in **Table 4**. In general, the influence of temperature on the interaction energies was greater than pressure impact. As for instance, $E_{VDW}$ between A3 and ortho-xylene was -439 KJ/mol at 300 K and 1 bar. Then it dropped 11% when the temperature was raised to 360 K. However, it increased by 6% when the pressure was boosted to 1000 bar. The influence of pressure and temperature changes on the number of HBs between model asphaltenes and water molecules was tested too. The results showed that the pressure change did not affect the number of HBs between asphaltenes and water molecules. On the contrary, the number of HBs increased slightly when the temperature was raised as reported in **Table 3**. For example, $(A2\text{-}W)^{Av. No. HBs}$ was 5.782 at 300 K and 1 bar. $(A2\text{-}W)^{Av. No. HBs}$ rose ~1% when the pressure was increased to 1000 bar and it decreased by ~8% when the temperature was raised to 360 K. Further, the effect of temperature and pressure on RDFs was studied (see **Figure 7**). RDFs confirmed the previous conclusions of temperature and pressure influence, i.e. pressure change did not affect $H_{COOH}$–$O_{water}$ in asphatene-water interaction when system condition changed from (300 K, 1 bar) to (300 K, 1000 bar) (see **Figures 7a** and **7b**). On the contrary, RDFs indicated that temperature had a noticeable impact on $H_{COOH}$–$O_{water}$ of asphatene-water interaction when the system condition changed from (300 K, 1 bar) to (360 K, 1 bar) (see **Figures 7a**, **7c**, and **7d**). These results confirmed that the strength of HBs was dropped as the temperature was raised consistent with the theory of hydrogen bonds.

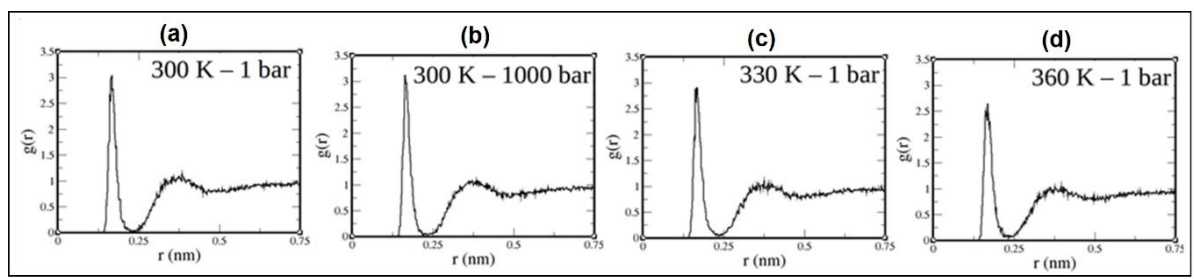

**Figure 7.** RDF of H$_{COO\underline{H}}$ - O$_{water}$ at various temperature and pressure.

# 4. Conclusions

In this paper, it is shown that polar functional groups of asphaltenes are responsible for hydrogen bonds generating with water molecules. The total interaction energies between asphaltene and ortho-xylene is much higher than between asphaltene and water. The large size of asphaltene produced strong van der Waals interaction between asphaltenes and solvents. Electrostatic interaction energies between asphaltenes and water were considerable while they were insignificant in the case of asphaltenes and ortho-xylene. The interaction energies decreased perceptibly when temperature was raised while it decreased comparatively when pressure was dropped.

**Acknowledgments:** We would like to thank J.R.M. Sagun for his technical assistance.